\documentclass[aps,prd,twocolumn,groupedaddress,showpacs,floatfix]{revtex4}

\usepackage{graphicx}
\graphicspath{{figures/}}

\begin{document}

\title{Tests of Lorentz violation in ${\bar\nu}_\mu  \rightarrow {\bar\nu}_e$ oscillations}
\author{
L.B.~Auerbach$^{10}$ 
R.L.~Burman$^7$, 
D.O.~Caldwell$^4$,
E.D.~Church$^2$,
A.K.~Cochran$^9$, 
J.B.~Donahue$^{7,}$\footnote{Deceased},
A.R.~Fazely$^9$,
G.T.~Garvey$^7$, 
R.~Gunasingha$^9$, 
R.L.~Imlay$^8$,
T.~Katori$^6$
W.C.~Louis$^7$, 
K.L.~McIlhany$^2$,
W.J.~Metcalf$^8$, 
G.B.~Mills$^7$,
V.D.~Sandberg$^7$, 
D.~Smith$^5$, 
I.~Stancu$^1$
W.H.~Strossman$^2$, 
R.~Tayloe$^6$,
M.~Sung$^8$, 
W.~Vernon$^3$,
D.H.~White$^7$,
and S.~Yellin$^4$ \\
\vspace{6pt}  
(LSND collaboration)}
\affiliation{
$^1$University of Alabama, Tuscaloosa, Alabama, 35487\\
$^2$University of California, Riverside,  California 92521\\
$^3$University of California, San Diego,  California 92093\\
$^4$University of California, Santa Barbara,  California 93106\\
$^5$Embry Riddle University, Prescott, Arizona 86301\\
$^6$Indiana University, Bloomington, Indiana 47405\\
$^7$Los Alamos National Laboratory, Los Alamos, New Mexico 87545\\
$^8$Louisiana State University, Baton Rouge, Louisiana 70803\\
$^9$Southern University, Baton Rouge, Louisiana 70813\\
$^{10}$Temple University, Philadelphia, Pennsylvania 19122}

\date{\today}

\begin{abstract}
A recently developed Standard-Model Extension (SME) formalism 
for neutrino oscillations that includes Lorentz and CPT violation 
is used to analyze the sidereal time variation of the neutrino event 
excess measured by the Liquid Scintillator Neutrino Detector (LSND) experiment.
The LSND experiment, performed at Los Alamos National Laboratory, 
observed an excess, consistent with neutrino oscillations, of ${\bar\nu}_e$ in a beam of  
${\bar\nu}_\mu$. It is determined that the 
LSND oscillation signal is consistent with no sidereal variation.  
However, there are several combinations of SME coefficients
that describe the LSND data; both with and without sidereal variations.
The scale of Lorentz and CPT violation extracted from the LSND data is of 
order $10^{-19}$~GeV for the SME coefficients $a_L$ and $E \times c_L$.
This solution for Lorentz and CPT violating neutrino oscillations
may be tested by other short baseline neutrino oscillation experiments, 
such as the MiniBooNE experiment.
\end{abstract}

\pacs{11.30.Cp, 14.60.Pq, 14.60.St}
\keywords{LSND, neutrino oscillations, Lorentz violation, and CPT violation}

\maketitle

\def\al{\alpha}
\def\be{\beta}
\def\ga{\gamma}
\def\de{\delta}
\def\ep{\epsilon}
\def\ve{\varepsilon}
\def\ze{\zeta}
\def\et{\eta}
\def\th{\theta}
\def\vt{\vartheta}
\def\io{\iota}
\def\ka{\kappa}
\def\la{\lambda}
\def\vpi{\varpi}
\def\rh{\rho}
\def\vr{\varrho}
\def\si{\sigma}
\def\vs{\varsigma}
\def\ta{\tau}
\def\up{\upsilon}
\def\ph{\phi}
\def\vp{\varphi}
\def\ch{\chi}
\def\ps{\psi}
\def\om{\omega}
\def\Ga{\Gamma}
\def\De{\Delta}
\def\Th{\Theta}
\def\La{\Lambda}
\def\Si{\Sigma}
\def\Up{\Upsilon}
\def\Ph{\Phi}
\def\Ps{\Psi}
\def\Om{\Omega}
\def\mn{{\mu\nu}}
\def\cA{{\cal A}}
\def\cl{{\cal L}}
\def\cE{{\cal E}}
\def\cI{{\cal I}}
\def\cN{{\cal N}}
\def\cS{{\cal S}}
\def\cO{{\cal O}}
\def\fr#1#2{{{#1} \over {#2}}}
\def\frac#1#2{\textstyle{{{#1} \over {#2}}}}
\def\pt#1{\phantom{#1}}
\def\prt{\partial}
\def\vev#1{\langle {#1}\rangle}
\def\ket#1{|{#1}\rangle}
\def\bra#1{\langle{#1}|}
\def\amp#1#2{\langle {#1}|{#2} \rangle}
\def\half{{\textstyle{1\over 2}}}
\def\lsim{\mathrel{\rlap{\lower4pt\hbox{\hskip1pt$\sim$}}
    \raise1pt\hbox{$<$}}}
\def\gsim{\mathrel{\rlap{\lower4pt\hbox{\hskip1pt$\sim$}}
    \raise1pt\hbox{$>$}}}
\def\Re{\hbox{Re}\,}
\def\Im{\hbox{Im}\,}
\def\Arg{\hbox{Arg}\,}
\def\etal {{\it et al.}}
\newcommand{\beq}{\begin{equation}}
\newcommand{\eeq}{\end{equation}}
\newcommand{\bea}{\begin{eqnarray}}
\newcommand{\eea}{\end{eqnarray}}
\newcommand{\bse}{\begin{subequations}}
\newcommand{\ese}{\end{subequations}}
\newcommand{\rf}[1]{(\ref{#1})}

\def\to{\rightarrow}
\def\from{\leftarrow}
\def\mix{\leftrightarrow}
\def\tofrom{\rightleftarrows}

\def\aL{(a_L)}
\def\cL{(c_L)}
\def\nub{\bar\nu}
\def\vp{\vec p}
\def\heff{h_{\rm eff}}
\def\As#1{({\cal A}_s)_{#1}}
\def\Ac#1{({\cal A}_c)_{#1}}
\def\Bs#1{({\cal B}_s)_{#1}}
\def\Bc#1{({\cal B}_c)_{#1}}
\def\C#1{({\cal C})_{#1}}
\def\Asa#1{({\cal A}_s^{(0)})_{#1}}
\def\Asc#1{({\cal A}_s^{(1)})_{#1}}
\def\Aca#1{({\cal A}_c^{(0)})_{#1}}
\def\Acc#1{({\cal A}_c^{(1)})_{#1}}
\def\Bsc#1{({\cal B}_s^{(1)})_{#1}}
\def\Bcc#1{({\cal B}_c^{(1)})_{#1}}
\def\Ca#1{({\cal C}^{(0)})_{#1}}
\def\Cc#1{({\cal C}^{(1)})_{#1}}
\def\nh^#1{{\hat N}^{#1}}
\def\indx{{\bar e\bar\mu}}

\section{Introduction}
Lorentz symmetry is one of the most fundamental ideas of both
relativistic local quantum field theory and general relativity. Early
tests, such as the Michelson-Morley and 
Kennedy-Thorndike experiments have established that Lorentz symmetry is
an exact symmetry of nature.  So it is natural to assume 
that Lorentz symmetry is an exact symmetry in the standard model (SM)
of particle physics. 
However, since the SM does not address gravity, a fundamental
theory of Planck-scale physics ($M_P \sim 10^{19}$~GeV),
including string theory~\cite{spontaneous}
and quantum gravity~\cite{Hawking},
may violate Lorentz and CPT symmetry~\cite{CPT04}.

If limited to conventional relativistic quantum
mechanics, it is possible to establish a self-consistent low-energy
effective theory with Lorentz and CPT violation; this is called the
standard-model Extension (SME)~\cite{Colladay}. 
The minimal-SME formalism has all the conventional
properties of the standard model including observer Lorentz
covariance, power counting renormalizability, energy momentum
conservation, quantized field, micro causality, and spin-statistics
with particle Lorentz and CPT violation due to background
Lorentz tensor fields of the universe. The
minimal SME also has $SU(3)_C \times SU(2)_L \times U(1)_Y$ gauge
invariance. Since the background Lorentz tensor fields are fixed in
spacetime, by definition, they do not transform under an active
transformation law. That implies rotation and boost dependence of physics
in a specific frame. This formalism focuses on the
inverse Planck-scale effect which is believed
to be suppressed by at least one order of the inverse Planck mass
($\sim {E \over {M_P}}$, where $E$ is the energy scale of the system under
consideration). Therefore, the physics quantities involved 
in the formalism are perturbative. 

Surprisingly, atomic physics has
achieved this sensitivity level, and extensive experimental studies have
been done (see, for example, Ref.~\cite{CPT04}). 
A recent experiment~\cite{Walsworth} of this type reaches a sensitivity 
to a specific combination of SME coefficients to order 
$\sim 10^{-32}$~GeV, well beyond a basic estimate of the scale 
of new physics.   In addition, spectral polarimetry of distant cosmological
sources yields a similar sensitivity for another combination of SME
coefficients~\cite{spectro}.
However, many of the SME coefficients still have
no experimental bounds.

Similarly, quantum interference experiments, such as meson
oscillations, are also sensitive to the small effect of Lorentz and CPT
violation~\cite{meson}. Tests have been made using data from many 
experiments, including KTeV~\cite{KTeV}, FOCUS~\cite{FOCUS}, BaBar~\cite{BaBar}, 
BELLE~\cite{BELLE}, and OPAL~\cite{OPAL}. 
Recently, the SME formalism for neutrino oscillations, another type of quantum
interference experiment, has become available~\cite{oscillation}.

\section {The LSND Evidence for Neutrino Oscillations}
The Liquid Scintillator Neutrino Detector (LSND) experiment~\cite{LSND2},
completed at the Los Alamos National Laboratory (LANL),
observed an excess of ${\bar\nu}_e$ in a beam of  
${\bar\nu}_\mu$ created from $\mu^{+}$ decay at rest.  The 
data analysis used the sample of detected ${\bar\nu}_e p \to e^+ n$
events with positron energy $20<E_{e^+}<60~\mathrm{MeV}$.
If interpreted as ${\bar\nu}_\mu$ to ${\bar\nu}_e$ oscillations,
this ${\bar\nu}_e$ excess implies a two-neutrino oscillation probability
of $(0.264 \pm 0.067 \pm 0.045) \%$. 
Here the first error is statistical and the second error is systematic
(neutrino flux, particle detection efficiency, cross sections, etc.). 
Despite the evidence 
for neutrino oscillations from
solar neutrinos~\cite{Homestake,kamiokande,GALLEX,SAGE,GNO,SNO},
atmospheric neutrinos~\cite{Super-K,MACRO}, accelerator neutrinos~\cite{K2K}, 
and reactor neutrinos~\cite{KamLAND}, the oscillation
signal observed at LSND remains a puzzle. 
Since the neutrino sector is thought as likely
to reveal new physics,
the LSND anomaly is often explained with new ideas
such as a mass-difference CPT-violating model 
(see, for example Ref.~\cite{Barenboim})
or sterile neutrino models 
(see Ref.~\cite{Janet} for a recent example). 
The MiniBooNE experiment~\cite{MiniBooNE} 
at Fermilab is currently taking data to test the LSND signal.

\section{Lorentz Violating Neutrino Oscillations}
Perhaps LSND is seeing the first signal of Planck-scale physics~\cite{TKRTCPT04}. 
To describe neutrino oscillations, including
Lorentz and CPT violation, a recently developed formalism for
neutrino oscillations~\cite{oscillation,SBL} using the SME 
framework~\cite{Colladay} is employed.  This framework allows for
a sidereal time variation of the neutrino oscillation probability.
 
Within the SME framework, the neutrino  free field Lagrangian becomes, 
\begin{eqnarray}
{\cal L} & = & 
{1 \over 2} i {\bar {\ps}}_A {\Ga}^{\mu}_{AB} \stackrel{\leftrightarrow}{D_{\mu}} {\ps}_{B} - {\bar {\ps}}_{A} M_{AB}  {\ps}_{B}, \\
\Ga^{\mu}_{AB} & = & 
{\ga}^{\mu} {\de}_{AB} + c^{\mu \nu}_{AB} {\ga}_{\nu} + d^{\mu \nu}_{AB} {\ga}_{5} {\ga}_{\nu} \nonumber \\
&   &
+ e^{\mu}_{AB} + i f^{\mu}_{AB} {\ga}_{5} + {1 \over 2} g^{\mu \nu \la}_{AB} {\si}_{\nu \la}, \\
M_{AB} & = & 
m_{AB} + i m_{5AB} \ga_5 \nonumber \\
&   & 
+ a^{\mu}_{AB} {\ga}_{\mu} + b^{\mu}_{AB} {\ga}_{5} {\ga}_{\mu} + {1 \over 2} H^{\mu \nu}_{AB} {\si}_{\mu \nu}.
\end{eqnarray}

The first term of $\Ga^{\mu}_{AB}$ and the first and second terms of
$M_{AB}$ are the only nonzero terms in the case of conventional neutrino
oscillations. The remaining terms in this Lagrangian represent
the physics of the background fields. In general, the background
Lorentz tensor fields are an infinite series, but if the focus is on a 
low-energy effective theory, these eight additional fields are complete. Here,
vacuum expectation values that contain $c^{\mu \nu}_{AB}$, $d^{\mu \nu}_{AB}$, and $H^{\mu\nu}_{AB}$ 
are CPT-even terms while $e^{\mu}_{AB}$, $f^{\mu}_{AB}$,
$g^{\mu \nu \la}_{AB}$, $a^{\mu}_{AB}$, and $b^{\mu}_{AB}$ are CPT-odd
by definition of the background fields. 
Notice that each background field
has flavor indices (A and B) that, unlike other systems, bring additional 
complication for the neutrino sector.

This Lagrangian leads to the modified Dirac equation,

\begin{equation}
(i\,\Ga^{\mu}_{AB} {\partial}_{\mu} - M_{AB}) {\ps}_B = 0.
\end{equation}

After some manipulation, this yields the effective Hamiltonian for active
neutrino oscillations~\cite{oscillation}. In particular, the effective
Hamiltonian for active antineutrino to antineutrino oscillations is,

\begin{eqnarray}
(h_{\mathrm{eff}})_{ab} 
& = & 
{| \vec p | {\de}_{ab} + \frac{({\tilde{m}}^2 )^{\ast}_{ab}}{2 | \vec p |}}
\nonumber \\
&  &
+ \frac{1}{| \vec p |}
[ - (a_L)^{\mu} p_{\mu} - (c_L)^{\mu \nu} p_{\mu} p_{\nu} ]^{\ast}_{ab}.
\label{eq:hamiltonian}
\end{eqnarray}

Here, the effective Hamiltonian is a $3 \times 3$ flavor Majorana basis
matrix of three active, right-handed, antineutrinos. The original effective
Hamiltonian~\cite{oscillation} can describe $\nu - \nu$, 
$\bar\nu - \bar\nu$, and $\nu-\bar\nu$ oscillations, but, in this work, 
lepton-number violating 
$\nu - \bar \nu$ oscillations are not considered. 
Therefore, the neutrino and antineutrino sectors can be diagonalized
separately. There is some coupling of SME coefficients,
$(a_L)^{\mu}_{ab} = (a)^{\mu}_{ab} + (b)^{\mu}_{ab}$ and $(c_L)^{\mu
\nu}_{ab} = (c)^{\mu \nu}_{ab} + (d)^{\mu \nu}_{ab}$. Also, other types
of SME coefficients do not show up in this analysis.
For the usual conventional neutrino oscillation case, the effective
Hamiltonian (Eq.~\ref{eq:hamiltonian}) contains only the first two terms. 
Then, the neutrino oscillation probability depends on  $\De m^2$ and the 
mixing matrix. But, in this general form, including
possible Lorentz and CPT violation, the diagonalization of the 
effective Hamiltonian is
more complicated and, in general, it can not be represented by $\De m^2$
and the mixing matrix alone.

\section{The Short-Baseline Approximation}
\label{sec:SBL}
If the baseline of the neutrino beam is short compared with the
neutrino oscillation length, $L$, the neutrino oscillation probability can be
expanded with an effective Hamiltonian. Expressed to leading 
order in $h_{\mathrm{eff}}$~\cite{SBL},

\begin{equation}
P_{{\bar \nu}_{\mu} \to {\bar \nu}_{e}} \simeq \frac{|(h_{\mathrm{eff}})_{\indx}|^2 L^2}{(\hbar c)^2}.
\label{eq:prob1}
\end{equation}

\begin{figure}
\includegraphics[width=\columnwidth]{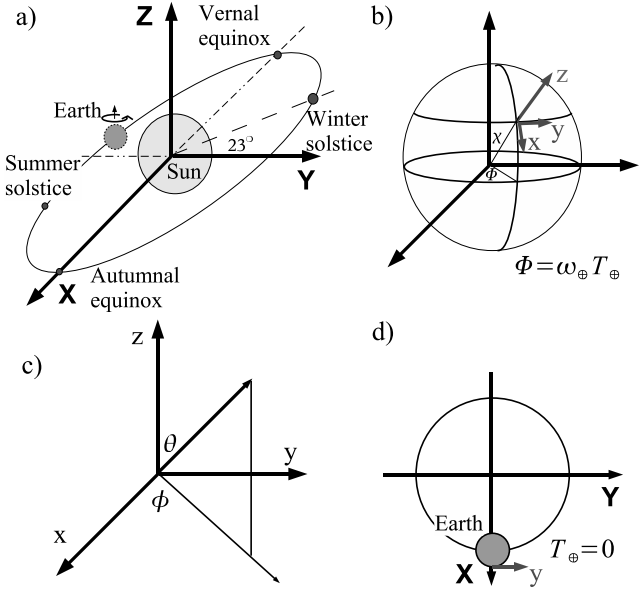}
\caption{\label{fig:coordinate}
Coordinate systems for the
sidereal time variation analysis: a) the Sun-centered system, b) the
Earth-centered system, c) the LANL local coordinate system, and d)
the definition of $T_{\oplus}=0$.}
\end{figure}

Note that in this equation, unlike the equations above, $\hbar$ and  $c$ have been 
explicitly included.
Since, in the effective Hamiltonian, $p^{\mu}$ contains information about
the propagation direction of the neutrino, this oscillation probability
depends on the neutrino propagation direction. 
In order to form a phenomenological expression for the
neutrino oscillation probability, it is most convenient to use 
a coordinate system fixed to the experiment~\cite{spectro,Bluhm}.
The standard choice is a
Sun-centered system (Fig.\ref{fig:coordinate}a) that is, to a good
approximation, an inertial frame for the experiment. 

With this choice of coordinates, the neutrino oscillation probability becomes,
\begin{eqnarray}
P_{\nub_\mu\to\nub_e} & \simeq & \fr{L^2}{(\hbar c)^2} |\, \C\indx 
  +\As\indx \sin\om_\oplus T_\oplus \nonumber \\
 & & 
  +\Ac\indx \cos\om_\oplus T_\oplus 
  +\Bs\indx \sin2\om_\oplus T_\oplus \nonumber \\
 & & 
  +\Bc\indx \cos2\om_\oplus T_\oplus \, |^2,
\label{eq:prob}
\end{eqnarray}
where
$\om_\oplus$ is the sidereal frequency (=2$\pi$/23h 56min 4.1s) and
$T_\oplus$ is the sidereal time as measured from a standard
origin.  Note that 
$P_{\nub_\mu\to\nub_e}$ may depend on the sidereal time. 

These parameters, $\C\indx$, $\As\indx$, $\Ac\indx$, $\Bs\indx$, and
$\Bc\indx$, depend on the SME coefficients $(a_L)^{\mu}$ and 
$(c_L)^{\mu\nu}$ and the
neutrino propagation direction unit vectors $\nh^X$,
$\nh^Y$, and $\nh^Z$ in the Sun-centered system. The direction
unit vectors depend on 
the colatitude $\ch$ of the experiment in the Earth-centered
system (Fig.\ref{fig:coordinate}b) and the zenith and azimuthal
angles $\th$ and $\ph$ of the ${\bar\nu}_\mu$ beam in the experiment 
local coordinate 
system (Fig.\ref{fig:coordinate}c).

\begin{equation}
\left(
\begin{array}{c}
\nh^X \\
\nh^Y \\
\nh^Z
\end{array}
\right) 
=
\left(
\begin{array}{c}
\cos \ch \sin \th \cos \ph + \sin \ch \cos \th \\
\sin \th \sin \ph \\
-\sin \ch \sin \th \cos \ph + \cos \ch \cos \th 
\end{array}
\right)
\end{equation}

For neutrinos from the Los Alamos Neutron Science Center
(LANSCE) beam to the LSND detector, $\ch = 54.1^{\circ}$, 
$\th = 99.0^{\circ}$ and $\ph = 82.6^{\circ}$~\cite{GPS}. 
The sidereal time is defined to be zero ($T_\oplus=0$)
at LANL local midnight on the autumnal equinox (Fig.\ref{fig:coordinate}d). 
At that time, the $y$ axis of the Earth-centered system coincides 
with the $Y$ axis of the Sun-centered system. The estimated error 
using this definition is three minutes, which is small 
compared to the time scale sensitivity of this
analysis.

Combining these values for neutrino propagation unit vectors
with the detailed expression of the parameters,
$\C\indx$, $\As\indx$, $\Ac\indx$, $\Bs\indx$, and $\Bc\indx$~\cite{SBL}, 
yields, for the particular case of the LSND
experiment:

\begin{eqnarray}
\C\indx 
&=& 
\frac{({\tilde{m}}^2 )_{\indx}}{2E}+[(a_L)^T_{\indx}+0.19(a_L)^Z_{\indx}]\nonumber \\
& &
+E[-1.48(c_L)^{TT}_{\indx}-0.39(c_L)^{TZ}_{\indx} 
\nonumber \\
& &
+0.44(c_L)^{ZZ}_{\indx}],
\label{eq:C}
\\
\As\indx 
&=& 
[0.98(a_L)^X_{\indx}+0.053(a_L)^Y_{\indx}] 
\nonumber\\
& &
+E[-1.96(c_L)^{TX}_{\indx}-0.11(c_L)^{TY}_{\indx} 
\nonumber\\
& &
-0.38(c_L)^{XZ}_{\indx}-0.021(c_L)^{YZ}_{\indx}],
\label{eq:As}
\\
\Ac\indx 
&=& 
[0.053(a_L)^X_{\indx}-0.98(a_L)^Y_{\indx}],
\nonumber\\
& &
+E[-0.11(c_L)^{TX}_{\indx}+1.96(c_L)^{TY}_{\indx}
\nonumber\\
& &
-0.021(c_L)^{XZ}_{\indx}+0.38(c_L)^{YZ}_{\indx}],
\label{eq:Ac}
\\
\Bs\indx 
&=& 
E[-0.052((c_L)^{XX}_{\indx}-(c_L)^{YY}_{\indx})
\nonumber \\
& &
+0.96(c_L)^{XY}_{\indx}],
\label{eq:Bs}
\\
\Bc\indx 
&=& 
E[0.48((c_L)^{XX}_{\indx}-(c_L)^{YY}_{\indx})
\nonumber \\
& &
+0.10(c_L)^{XY}_{\indx}].
\label{eq:Bc}
\end{eqnarray}

In Eq.~\ref{eq:C}, the mass-squared term, 
${\tilde{m}}^2_{\indx}$, has been included.  
This allows for conventional 
massive-neutrino oscillations in addition to the 
Lorentz-violation oscillations.
It is assumed that the size of this term does 
not invalidate the short-baseline approximation~\cite{SBL}.

\section{Sidereal Time Distribution of the LSND Data}
\label{sc:sidetime}
In conventional explanations of neutrino oscillations, the oscillation
probability is independent of sidereal time and, therefore, the
sidereal time distribution of oscillation events is expected to be 
constant. 
In the Lorentz and CPT violating
model of neutrino oscillations considered here, nonzero values of the
model parameters could exhibit themselves as modulations to the
sidereal time distribution (as in Eq.~\ref{eq:prob}). 
The sidereal time dependence of candidate oscillation events
from the LSND data sample has been examined and 
subjected to statistical tests
to quantify any evidence for a sidereal variation.  

In the analysis of the final LSND data set~\cite{LSND2}, 205 neutrino oscillation
candidate events were reported with positron energy in the range 
$20 < E_{e^+} < 60~\mathrm{MeV}$ and 
with an identified neutron-capture photon.
There are two classes of background in the oscillation
sample: beam-unrelated and beam-related ($\nu$-induced).
The beam-unrelated backgrounds arise from cosmic ray processes. It
is measured in beam-off data and then subtracted from the beam-on data.
The beam-related backgrounds are calculated from known neutrino (nonoscillation) 
interactions.     

The neutrino beam used for the LSND experiment was produced using 
protons from the LANSCE accelerator~\cite{LSND-NIM}. The proton beam 
was delivered at approximately 100Hz in pulses of 600 $\mu$s duration.
The detector was triggered independently of the state of the beam and
the beam status was recorded.  In this manner,
beam-off data was taken continuously in the time between beam pulses.
The resulting beam-off data set was approximately 16 times larger
than the beam-on data set.
This allowed for an accurate measurement of the 
beam-unrelated background by weighting 
the beam-off data by the beam duty-factor (calculated for
each run).  

The estimated number of beam-unrelated and $\nu$-induced
background events in this sample are $106.8\pm2.5$ and $39.2\pm3.1$,
respectively. 
These events were collected during experimental running in 1993 through
1998.  There were six sets of runs, one in each of these years.  The GPS 
(Global Positioning System) time stamp, necessary for this analysis, was
not included into the LSND data stream until midway through the 1994
run period. Because of this, only 186 of the 205 oscillation candidate
events could be used in this analysis.  The expected numbers of
beam-off and $\nu$-induced backgrounds in this smaller sample are $94.0 \pm 2.3$ and
$35.6 \pm 2.8$, respectively.

\begin{figure}
\includegraphics[width=\columnwidth]{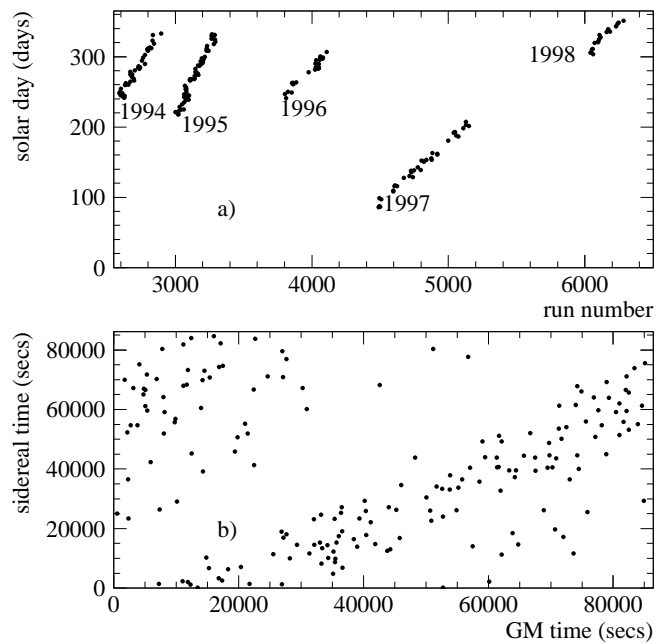}
\caption{\label{fig:day} 
Distribution of beam-on neutrino candidate
events in a) run number vs. solar day (day 1= January 1) and b) GM
time vs sidereal time. The year of each set of runs is indicated in~a).}
\end{figure}

Ideally, an experiment to search for sidereal variations in a signal
would run continuously throughout the calendar year so that one
particular sidereal time bin would be drawn from the entire range of
local time.  This was not the case with LSND, but runs did 
cover the space of local time vs sidereal time with reasonable 
completeness, as can be seen in Figure~\ref{fig:day}. The 
Los Alamos (clock) time can be determined from Greenwich Mean (GM) 
time by subtracting 6 (7) hours in the summer (winter).

To quantify the statistical significance of any sidereal time 
variation in the data, we employed two different statistical tests:
a Pearson's-$\chi^2$ test~\cite{Frodesen,PDG} and a
Kolmogorov-Smirnov (KS) test~\cite{Frodesen}.  In both of these
tests, the data were compared to the (null) hypothesis that 
the event rate is constant in sidereal time.
Note that this null hypothesis is not that  
no oscillation signal exists, but only that the signal is constant
in time.
We also examined the GM time distributions and 
applied these tests with the null hypothesis of an underlying
distribution that is constant in GM time.   

 
The Pearson's-$\chi^2$ (P-$\chi^2$), as implemented in this analysis,
is
\begin{equation}
\label{eq:Pchi2}
\mathrm{P-}\chi^2=\sum_{i=1}^N \frac{(n_i-\nu_i)^2}{\nu_i}; \nu_i=n/N,  
\end{equation}
where $n$ is the total number of events in the sample, $N$ is the number 
of time bins, and $n_i$ is the measured number of events 
in time bin $i$.  The predicted number of events in each time bin, $\nu_i$, 
is constant for each time bin.  Note that this quantity is constructed
with the variance of the {\em expected} number of events in the denominator.

The P-$\chi^2$ statistic will follow, in the absence of sidereal time
variations and with sufficient events per time bin, a $\chi^2$ distribution 
with number of degrees of freedom equal to the number of time bins 
minus one~\cite{Frodesen,PDG}. The standard criterion for sufficient
events is that $\nu_i \ge 5$~\cite{Frodesen,PDG}.  The binning for
the beam-on data has been chosen to satisfy this.
The $p$-value, ($P(\chi^2$), one minus the $\chi^2$ cumulative distribution) 
can be extracted and interpreted as a confidence level that the
null hypothesis explains the data.

The KS test has the advantage that it works with unbinned data,
thus eliminating the need to choose a binning.  It
involves a comparison between the data and the null hypothesis
via cumulative distributions.  
Unlike the  P-$\chi^2$ test, 
it is sensitive to ``runs''
in the data, thus making the P-$\chi^2$ and KS tests complementary.
The KS statistics reported here are the maximum cumulative 
deviation, $D_n$, and the KS probability, $P(\mathrm{KS})$.  The quantity
$P(\mathrm{KS})$, which is obtained from the known distribution of $D_n$~\cite{Frodesen}, 
can be interpreted as a confidence level that the data is explained by the null
hypothesis. 

\begin{figure}
\includegraphics[width=\columnwidth]{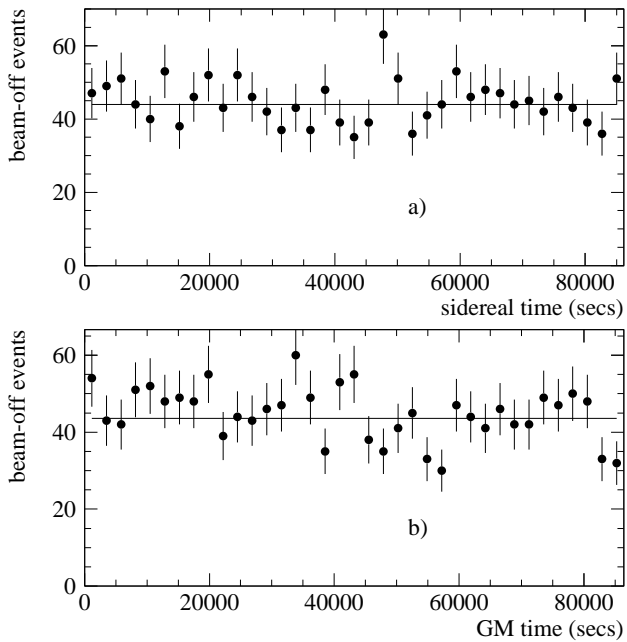}
\caption{\label{fig:boff}
Sidereal a) and GM b) time distributions of beam-off
data using 37 time bins.
The solid lines indicate the expected underlying distribution with no
sidereal time variation.  The errors bars displayed in this plot 
(and subsequent) are the square root of the number in each bin.}
\end{figure}

In the LSND data set considered for this analysis, there were
1656 beam-off events passing the neutrino oscillation cuts.  These events,
after weighting for the beam-on duty-factor, determine the number
and distributions of beam-unrelated background events in the beam-on sample.
The distribution of sidereal and GM times in 37 time bins for these beam-off 
events is shown in Fig.~\ref{fig:boff}. 
The number of time bins chosen for this distribution 
was obtained by applying the $N>5$ criterion for the beam-on data.
The errors shown in Fig.~\ref{fig:boff} (and subsequent figures)
are the square root of the number of counts in the bin.  Note that these 
errors are not used in the calculation of the P-$\chi^2$ (Eq.~\ref{eq:Pchi2}).
The P-$\chi^2$ is 29.6 for 37 sidereal time bins corresponding
to $P(\chi^2) = 0.77$.  The KS test on this same data yields
$D_n=0.019$ and $P(\mathrm{KS}) = 0.60$. These results indicate that the
beam-off data are in reasonable agreement with the null hypothesis
(no sidereal time dependence).  The GM time distribution yields a slightly
low $P(\mathrm{KS})=0.01$, however, for this same distribution $P(\chi^2)=0.29$.  
In addition, any GM time variations are distributed throughout
a range in sidereal time.  For these reasons, we conclude that there 
is no evidence for substantial environmental or ``day-night'' 
sidereal variations in the beam-unrelated backgrounds.

\begin{figure}
\includegraphics[width=\columnwidth]{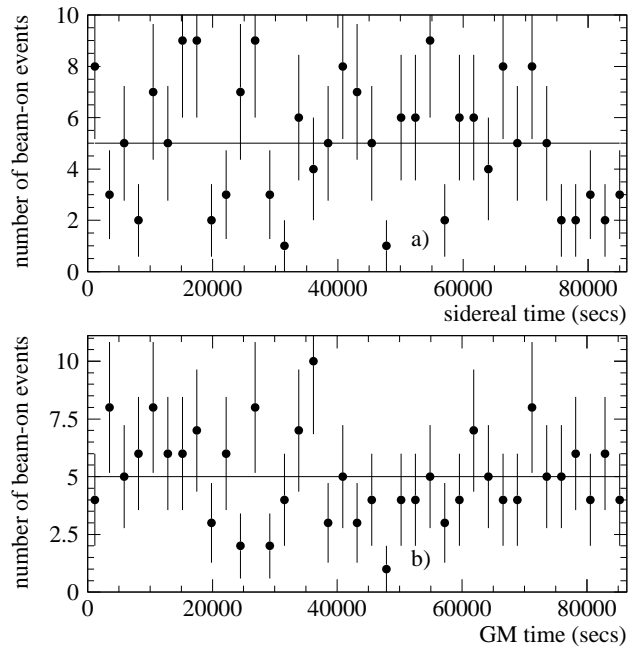}
\caption{\label{fig:bon} 
Sidereal a) and GM b) time distributions of
the 186 beam-on oscillation candidate events in 37 bins.  The
solid lines indicate the expected underlying distribution with no
sidereal time variation.}
\end{figure}

The sidereal and GM time distributions of the 186 oscillation candidate 
events are shown in Figure~\ref{fig:bon}.  The P-$\chi^2$ for
the sidereal time distributions is 44.8 for 
37 time bins. The corresponding $p$-value for the 
sidereal time distribution is $P(\chi^2)=0.15 $.  
A KS test applied to these distributions yields $P(\mathrm{KS})=0.234$.
The sidereal time distribution is slightly less compatible with
no time variation as is evident in both of these statistical tests.  
However, the variation is not statistically significant.
A KS test between beam-on
and beam-off data was also applied and shows compatibility between 
the two data sets. 
The complete results from the statistical
tests on the sidereal and GM time distributions for beam-on and beam-off 
data are summarized in Table~\ref{tab:statcomp}.

\begin{table}
\begin{center}
\begin{tabular}{|c|cc|cc|}
\hline
\multicolumn{5}{|c|}{null hypothesis tests} \\
\hline \hline
 & \multicolumn{2}{c|}{beam-on} &  \multicolumn{2}{c|}{beam-off} \\
 & sidereal & GM & sidereal & GM \\
\hline
\# of events & \multicolumn{2}{c|}{186} & \multicolumn{2}{c|}{1656} \\
\hline
\multicolumn{5}{|l|}{\bf Pearson's $\chi^2$:} \\
\hline
$N_\mathrm{bins}$ & 37 & 37 & 37 & 37 \\
$\chi^2$ & 44.8 & 27.6 & 29.6 & 40.3 \\
$P(\chi^2)$ & 0.15 & 0.84 & 0.77 & 0.29 \\
\hline
\multicolumn{5}{|l|}{\bf Kolmogorov-Smirnov:} \\
\hline
$D_n$ & 0.076 & 0.066 & 0.019 & 0.040 \\
$P(\mathrm{KS})$ & 0.234	& 0.386	& 0.604	& 0.010 \\
\hline 
\multicolumn{5}{c}{}
\end{tabular}

\begin{tabular}{|c|cc|}
\hline 
\multicolumn{3}{|c|}{beam-on/beam-off tests} \\
\hline \hline
 & sidereal & GM \\
\hline 
$D_n$ & 0.067 &	0.046 \\ 
$P(\mathrm{KS})$ &  0.432 & 0.864 \\ 
\hline
\end{tabular}

\end{center}
\caption{A summary of results from statistical tests on the
sidereal and GM time distributions of the $20<E_{e^+}<60$~MeV 
neutrino oscillation data. The null hypothesis tests compare
the data with a constant time distribution.  The beam-on/beam-off
tests compare the two sets.}
\label{tab:statcomp}
\end{table}

\begin{figure}
\includegraphics[width=\columnwidth]{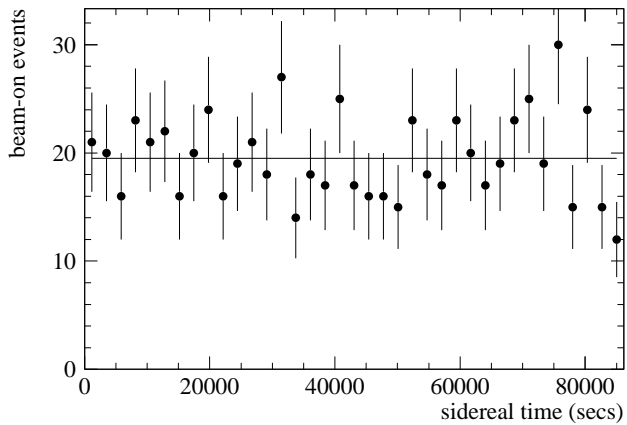}
\caption{\label{fig:nueccheck} 
Sidereal time distribution of beam-on data with 
$^{12}C(\nu_e,e^-)^{12}N_{\mathrm{g.s.}}$ cuts.
The line indicates the average value.}
\end{figure}

To check the underlying assumption that the beam was delivered with
equal efficiency throughout the sidereal day, a sample of 
$^{12}C(\nu_e,e^-)^{12}N_{\mathrm{g.s.}}$ events was obtained by 
applying cuts to select for subsequent $\beta$-decays of 
$^{12}N_{\mathrm{g.s.}}$ (as described in Ref.~\cite{LSND-nueC}).
This procedure yields 722 beam-on events with a beam-unrelated 
background of 17.5 events.  The sidereal time distribution of these
beam-on events are shown in Fig.~\ref{fig:nueccheck}.  The P-$\chi^2$ 
for this sidereal time distributions is 29.3 for 37 time bins
which corresponds to $P(\chi^2)=0.78$.  The Kolmogorov-Smirnov
test yields $D_n=0.020$ and $P(\mathrm{KS}) = 0.94$.
The values indicate that the assumption of constant
beam delivery, averaged over the sidereal day and over all LSND runs,
is consistent with the data.

\section{The Extraction of the SME Parameters}
\label{sc:SMEextract}
While the LSND oscillation data examined in the previous section shows
no statistically significant sidereal time variation, it is interesting
to examine the data in context of the SME model explained in Section~\ref{sec:SBL}.
First, this model does not require a sidereal time variation and, second, 
the LSND data set does allow for some sidereal time variation.     

A maximum-likelihood method, with Eq.~\ref{eq:prob} as a description of 
the oscillation signal, was performed to extract allowed values of the SME
parameters, $\C\indx$, $\As\indx$, $\Ac\indx$, $\Bs\indx$, and $\Bc\indx$.
In general, these parameters are complex ---  
the special case is considered here where these parameters are real. Also, the
values extracted are an effective average over the energy 
range of the LSND data set, $20 < E_{e^+} < 60~\mathrm{MeV}$.

The parameters were extracted using an 
unbinned likelihood function,
\begin{eqnarray}
\Lambda &=& {{e^{-\mu}} \over {N!}} \prod_{i=1}^{N} (\mu_s \mathcal{F}_s + \mu_b \mathcal{F}_b) 
\nonumber \\
& &
\times \prod_{i=\mathrm{s,b}}
{1 \over {\sqrt{2\pi {\si_i}^2}}}\exp \left( -{{(\mu_i - \bar {\mu_i})}^2 \over {2 {\si_i}^2}} \right) 
\end{eqnarray}
where $N$ is the total number of events in the sample, $\mu_s$ is the
total predicted oscillation signal events, $\mu_b$ is the estimated
number of background events, and $\mu = \mu_s + \mu_b$.  The shape of
the data in sidereal time is described with the functions
$\mathcal{F}_s$ and $\mathcal{F}_b$. $\mathcal{F}_s$ depends on the
SME parameters as in Eq. \ref{eq:prob} and $\mathcal{F}_b$ is assumed
to be constant in sidereal time. The latter half of the likelihood 
function describes systematic errors on the signal and background events. 
In implementation, the natural log of
the likelihood function, $\ell$ $(=\ln \Lambda)$, was used. Note
that this function describes both the shape and the overall number of
events.

Three different parameter combinations were considered.
\begin{itemize}
\item{\bf 1-parameter:} \\
\nopagebreak
$\C\indx \neq 0; \As\indx,\Ac\indx,\Bs\indx,\Bc\indx = 0$  \\
The ``rotationally invariant'' 
case~\cite{Coleman,Pakvasa,Bahcall,Pena-Garay}.
\item{\bf 3-parameter:} \\
\nopagebreak
$\C\indx,\As\indx,\Ac\indx \neq 0; \Bs\indx,\Bc\indx = 0$ \\
Includes all of the CPT-odd terms of the minimal-SME model. 
\item{\bf 5-parameter:} \\
\nopagebreak
$\C\indx,\As\indx,\Ac\indx,\Bs\indx,\Bc\indx \neq 0$ \\
Full minimal-SME model including 
both CPT-odd and CPT-even terms. 
\end{itemize}

\begin{figure}
\includegraphics[width=\columnwidth]{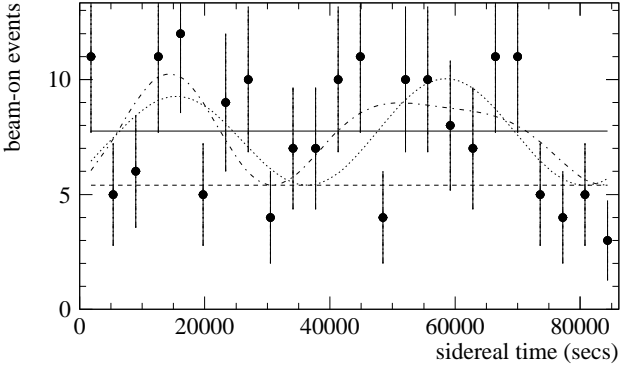}
\caption{\label{fig:parfit}
Sidereal time distribution of the LSND oscillation data
in 24 time bins together with the maximum-$\ell$ 
solutions for the 1-parameter (solid line), 
3-parameter (dotted), and 5-parameter (dot-dashed) combinations.
The dashed line indicates the estimated background contribution.}
\end{figure}

Using each of these three parameter sets, the log likelihood,
$\ell$, was calculated for the 186 candidate oscillation events
as each of the parameters in the set was
varied in a range around zero. The sidereal time for the
parameter values that maximized $\ell$ 
is plotted together with the data in Figure~\ref{fig:parfit}.
Note that the data in Fig.~\ref{fig:parfit} is grouped into 24 time
bins instead of 37 as was used in Fig.~\ref{fig:bon}.  This is
to allow for the quality of the fit to be more easily seen.
 
\begin{figure}
\includegraphics[width=\columnwidth]{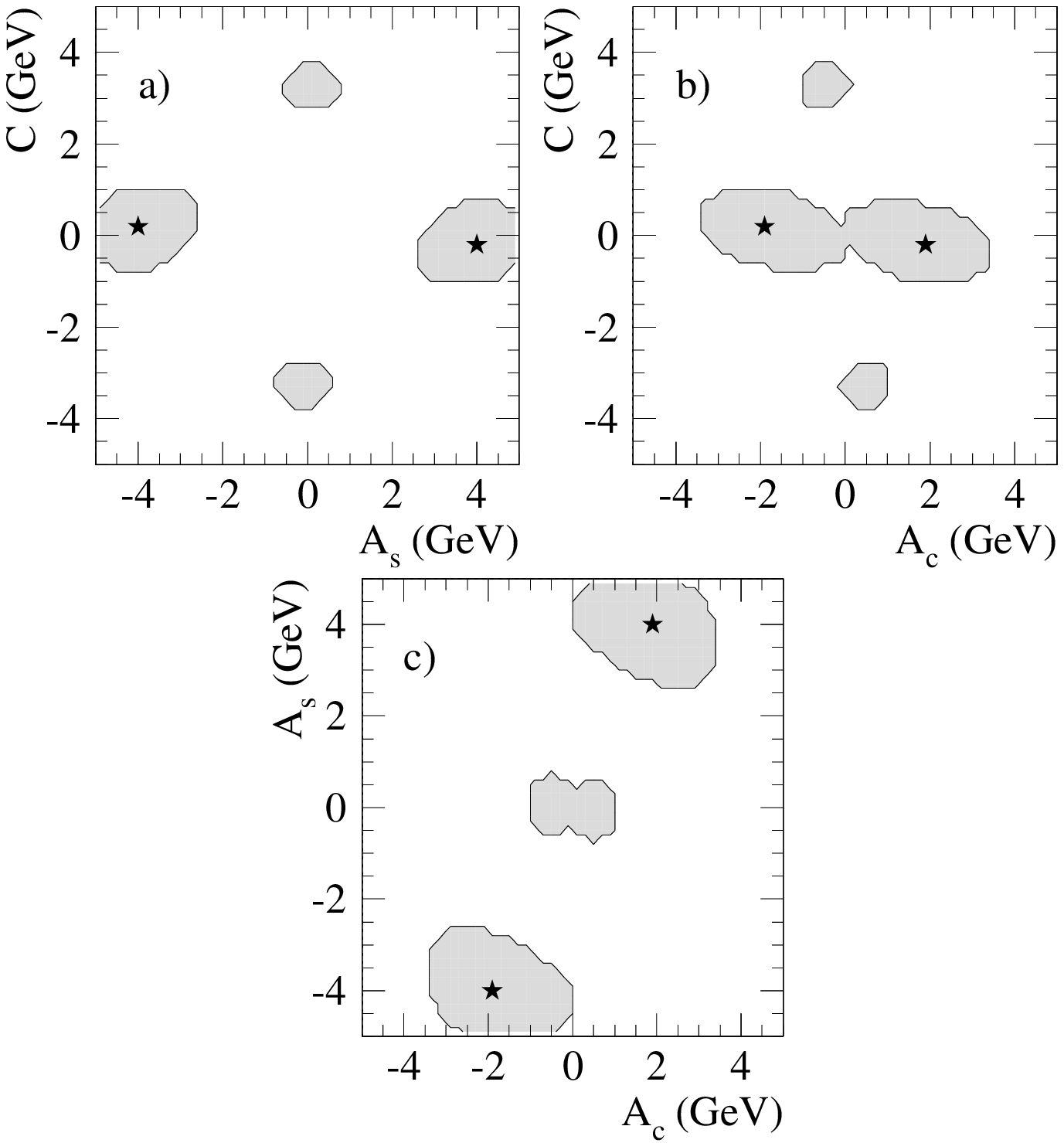}
\caption{\label{fig:contour3}
Log likelihood values for the 3-parameter description of 
the LSND sidereal time distribution: a) $\As\indx$ vs $\C\indx$, b) $\Ac\indx$ vs $\C\indx$,
and c) $\Ac\indx$ vs $\As\indx$. 
The contours in a)-c) indicate the 1-$\sigma$ (total error) 
allowed regions ($\ell >
\ell_\textrm{max}-1.77$) and the stars indicate the maximum-$\ell$
parameter values.}
\end{figure}

The maximum-$\ell$ solutions are summarized below.  
The likelihood contours for the 3-parameter combination are shown in Fig.~\ref{fig:contour3}.
The (1$\sigma$)  errors were
calculated by determining the parameter ranges where $\ell$
decreased by 0.5 (1-parameter), 1.77 (3-parameters), or 3.0
(5-parameters) from the maximum value. 

\begin{itemize}
\item{\bf 1-parameter:} 
\begin{eqnarray}
\C\indx & = & 3.3\pm0.4\pm0.2
\end{eqnarray}
\item{\bf 3-parameter:} \\
There are two solutions within the $1\sigma$
likelihood region (see Fig.~\ref{fig:contour3}). \\
Solution 1 (maximum-$\ell$): 
\begin{eqnarray}
\C\indx  & = & -0.2\pm1.0\pm0.3, \nonumber \\
\As\indx & = & 4.0\pm1.3\pm0.4,  \nonumber \\
\Ac\indx & = & 1.9\pm1.8\pm0.4. 
\end{eqnarray}
Solution 2: 
\begin{eqnarray}
\C\indx  & = & 3.3\pm0.5\pm0.3, \nonumber \\ 
\As\indx & = & 0.1\pm0.6\pm0.2, \nonumber \\
\Ac\indx & = & -0.5\pm0.6\pm0.2.
\end{eqnarray}
\item{\bf 5-parameter:} Multiple (connected) solutions exist in the 5-parameter
case making a numerical extraction of errors impossible.  
The maximum-$\ell$ solution is: 
\begin{eqnarray}
\C\indx  & = & -0.7, \nonumber \\ 
\As\indx & = & 3.7,  \nonumber \\
\Ac\indx & = & 2.3,  \nonumber \\
\Bs\indx & = & 0.9,  \nonumber \\ 
\Bc\indx & = & -0.6. 
\end{eqnarray}
\end{itemize}
All parameters have units of $10^{-19}$~GeV and the errors quoted above 
are in the form $\pm$(statistical)$\pm$(systematic).

In all of these results, duplicate solutions exist with opposite
signs for all of the parameters. Note that in the 3-parameter case,
the two solutions correspond to a large value for $\C\indx$ with a small
value for $\As\indx,\Ac\indx$ and {\em vice versa}.  The small-$\C\indx$, 
large-$\As\indx,\Ac\indx$ solution is only slightly favored over the 
large-$\C\indx$, small-$\As\indx,\Ac\indx$ solution.  This is because the
sinusoidal terms in Eq.~\ref{eq:prob} improve the description
of the data in sidereal time, although, an oscillation probability 
that is constant in sidereal time is consistent with the data (as 
was reported in Sec.~\ref{sc:sidetime}).  Note also that the solution
where $\C\indx$ is the only nonzero term can be identified with
the conventional neutrino oscillation description via the first term
of Eq.~\ref{eq:C}. A solution
with all parameters $\approx0$ is highly disfavored. This is
equivalent to the statement that the LSND oscillation excess is
statistically significant.

Since the oscillation probability depends on the SME parameters
{\em squared}, the results for the SME parameters obtained
above are more easily compared to the measured oscillation probability 
from LSND via combinations of
the squares of the parameters.  The value resulting from the 1-parameter 
solution is
\begin{equation}
|\C\indx|^2 = 10.7 \pm 2.6 \pm 1.3~(10^{-19}\textrm{GeV})^2.
\end{equation}
The values for the parameter square sum resulting from the multiparameter
combinations are more highly constrained 
than for individual parameters.  The value
extracted from the 3-parameter solution is
\begin{eqnarray}
|\C\indx|^2+\half|\As\indx|^2+\half|\Ac\indx|^2 \;\;\;\;\; & & \nonumber \\ 
\;\;\;\;\; = 9.9\pm2.3\pm1.4~(10^{-19}\textrm{GeV})^2, & &  
\end{eqnarray}
and from the 5-parameter solution,
\begin{eqnarray}
|\C\indx|^2+ \half|\As\indx|^2+\half|\Ac\indx|^2  \;\;\;\;\ & & \nonumber \\
\;\;\;\;\ + \half|\Bs\indx|^2+\half|\Bc\indx|^2 & &  \nonumber \\
\;\;\;\;\ = 10.5\pm2.4\pm1.4~(10^{-19}\textrm{GeV})^2. && 
\end{eqnarray}
These results for the combination of SME parameters are consistent
with the previously reported oscillation probability from LSND~\cite{LSND2}
and with the estimate presented in Ref.~\cite{SBL}.

\section{A High-Energy Subset of the LSND Data}
A high-energy subset of the LSND data with a positron energy cut, 
$36 < E_{e^+} < 60~\mathrm{MeV}$,
is interesting to examine separately. 
The $\nu$-induced background is
reduced in this sample~\cite{LSND2}.  Furthermore, the $1/E$ prefactor to 
the mass term in Eq.~\ref{eq:C} would suppress the conventional oscillation
terms relative to any Lorentz-violation terms present.

\begin{figure}
\includegraphics[width=\columnwidth]{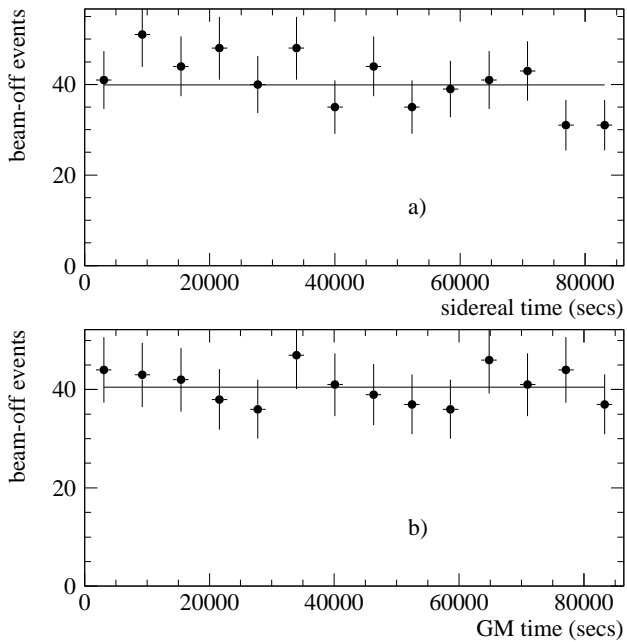}
\caption{\label{fig:hiEboff} 
Sidereal a) and GM b) time distributions of the high-energy, beam-off data
in 14 bins.
The solid lines indicate the expected underlying distribution with no
sidereal time variation.}
\end{figure}

This reduced data set consists of 73 beam-on events with expected
beam-unrelated and $\nu$-induced
background events of $31.3\pm0.8$ and $10.0\pm0.8$,
respectively.  The sidereal and GM time distributions of the 571 beam-off
events passing these high-energy cuts are shown in 
Figure~\ref{fig:hiEboff}. 
The P-$\chi^2$ is 12.2 for 14 sidereal time bins, corresponding
to $P(\chi^2) = 0.51$.  The resulting $p$-value from the KS test to
this distribution is $P(\mathrm{KS}) = 0.080$.  Again, these values show 
no reason to reject the null hypothesis for beam-off data.  

\begin{figure}
\includegraphics[width=\columnwidth]{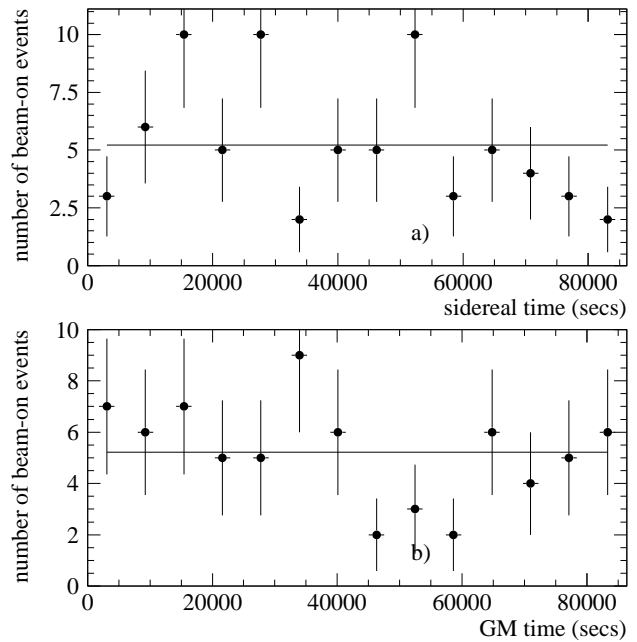}
\caption{\label{fig:hiEbon}
Sidereal a) and GM b) time distributions of
the 73 LSND high-energy, beam-on oscillation candidate events in 14 bins.  The
solid lines indicate the expected underlying distribution with no
sidereal time variation.}
\end{figure}

The sidereal time distribution of the high-energy beam-on data
is shown in Figure~\ref{fig:hiEbon}.  The P-$\chi^2$ 
is 20.4 for 14 sidereal time bins corresponding
to $P(\chi^2) = 0.09$.  The resulting $p$-value for the KS test is
$P(\mathrm{KS}) = 0.178$.  Although these values indicate a slightly reduced
agreement with the null hypothesis, they do not indicate a
statistically significant sidereal variation.
The complete results from the statistical
tests on the sidereal and GM time distributions for the 
high-energy data are summarized in Table~\ref{tab:hiEstatcomp}.

\begin{table}
\begin{center}
\begin{tabular}{|c|cc|cc|}
\hline
\multicolumn{5}{|c|}{null hypothesis tests} \\
\hline \hline
 & \multicolumn{2}{c|}{beam-on} &  \multicolumn{2}{c|}{beam-off} \\
 & sidereal & GM & sidereal & GM \\
\hline
\# of events & \multicolumn{2}{c|}{73} & \multicolumn{2}{c|}{571} \\
\hline
\multicolumn{5}{|l|}{\bf Pearson's $\chi^2$:} \\
\hline
$N_\mathrm{bins}$ & 14 & 14 & 14 & 14 \\ 
$\chi^2$ & 20.4 & 9.7 & 12.2 & 	4.4 \\
$P(\chi^2)$ & 0.09 & 0.72 & 0.51 & 0.99 \\
\hline
\multicolumn{5}{|l|}{\bf Kolmogorov-Smirnov:} \\
\hline
$D_n$ & 0.129 & 0.123 & 0.053 & 0.026 \\
$P(\mathrm{KS})$ & 0.178	& 0.221	& 0.080	& 0.826 \\ 
\hline 
\multicolumn{5}{c}{}
\end{tabular}

\begin{tabular}{|c|cc|}
\hline 
\multicolumn{3}{|c|}{beam-on/beam-off tests} \\
\hline \hline
 & sidereal & GM \\
\hline 
$D_n$ & 0.094 &	0.107 \\
$P(\mathrm{KS})$ & 0.621	& 0.451 \\
\hline
\end{tabular}

\end{center}
\caption{A summary of results from statistical tests on the
sidereal and GM time distributions of the LSND $36<E_{e^+}<60$~MeV 
neutrino oscillation data.  These values result from the
same procedure as used for Table~\ref{tab:statcomp}.}
\label{tab:hiEstatcomp}
\end{table}

\begin{figure}
\includegraphics[width=\columnwidth]{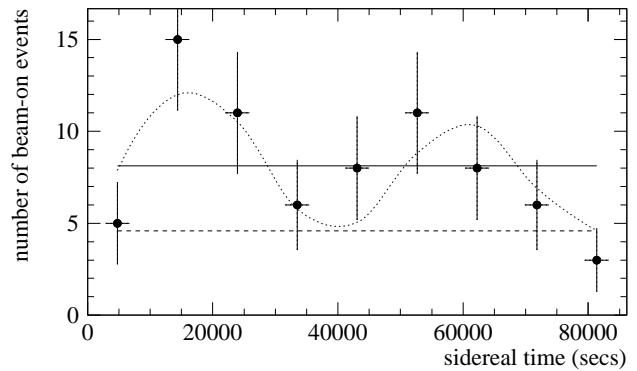}
\caption{\label{fig:hiEparfit}
Sidereal time distribution of the high-energy
LSND oscillation data
in 9 time bins together with the maximum-$\ell$ 
solutions for the 1-parameter (solid line) and 
3-parameter (dotted) combinations.
The dashed line indicates the estimated background contribution.}
\end{figure}

The maximum-likelihood procedure was applied to this high-energy
data set using the 1- and 3-parameter combinations described in
Section~\ref{sc:SMEextract}.  The limited data sample did not
allow for the 5-parameter combination.
Figure~\ref{fig:hiEparfit} (with a reduced bin size) 
shows 1- and 3-parameter maximum-$\ell$
solutions superimposed on the high-energy data.
Both parameter combinations produce acceptable descriptions of 
the data.

\begin{figure}
\includegraphics[width=\columnwidth]{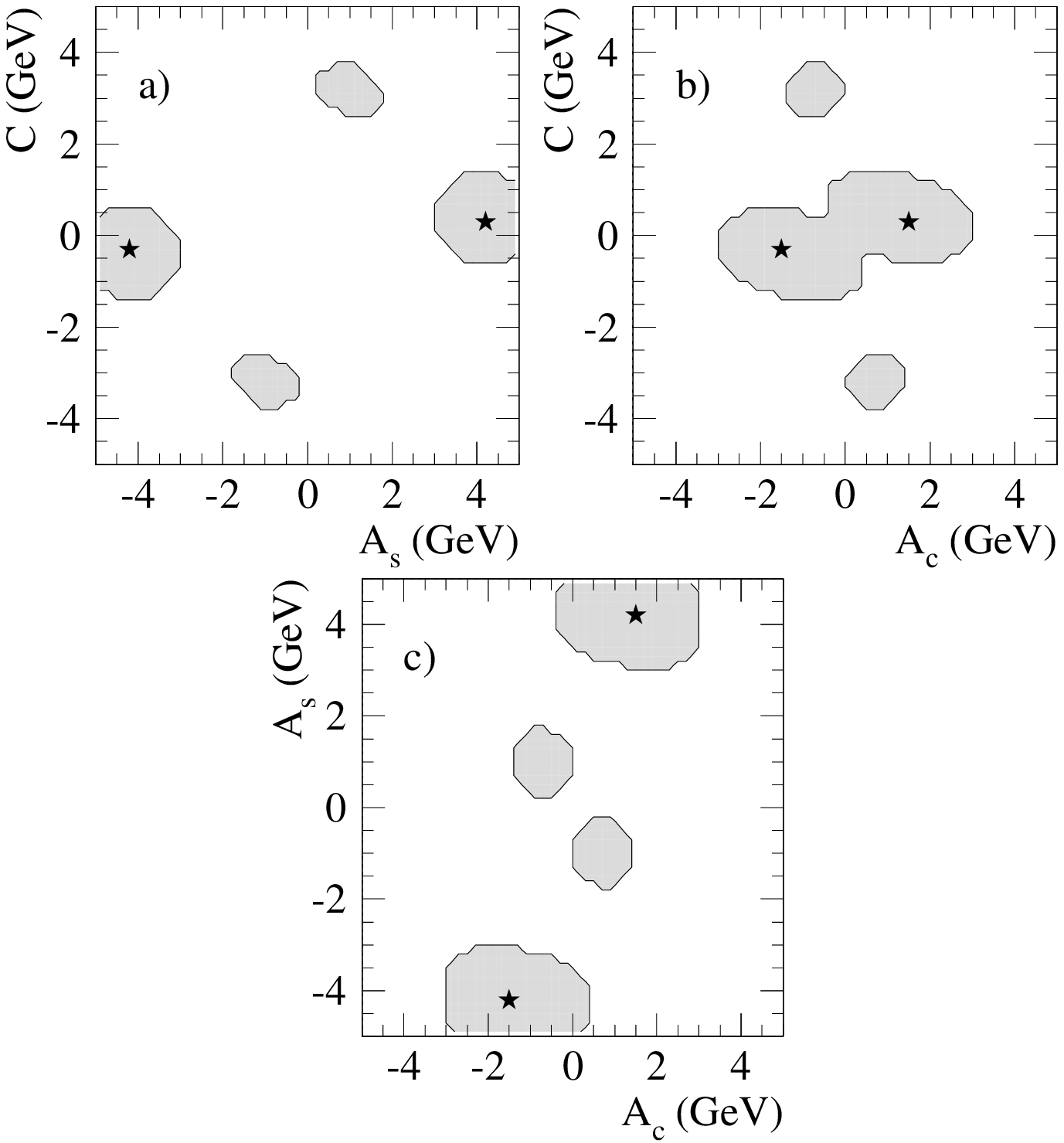}
\caption{\label{fig:hiEcontour3}
Log likelihood values for the 3-parameter description of 
the high-energy LSND sidereal time distribution: a) $\As\indx$ vs $\C\indx$, b) $\Ac\indx$ vs $\C\indx$,
and c) $\Ac\indx$ vs $\As\indx$. 
The contours in a)-c) indicate the 1-$\sigma$ (total error) allowed regions ($\ell >
\ell_\textrm{max}-1.77$) and the stars indicate the maximum-$\ell$
parameter values.}
\end{figure}

The values for the parameter square sums extracted with the 
maximum-likelihood method are summarized below.
The likelihood contours for the 3-parameter combination are shown 
in Fig.~\ref{fig:hiEcontour3}.

\begin{itemize}
\item{\bf 1-parameter:}
\begin{eqnarray}
|\C\indx|^2 & = & 10.7 \pm 2.9 \pm 1.5~(10^{-19}\textrm{GeV})^2
\end{eqnarray}
\item{\bf 3-parameter:} 
\begin{eqnarray}
|\C\indx|^2+\half|\As\indx|^2+\half|\Ac\indx|^2 \;\;\;\;\; & & \nonumber \\ 
\;\;\;\;\;\;\; = 10.2\pm2.7\pm1.3~(10^{-19}\textrm{GeV})^2 & & 
\end{eqnarray}
\end{itemize}

As can be seen by comparing the results from the high-energy
subset with the entire data set, there are no significant differences.
The time distributions from the high-energy subset
are consistent with no sidereal variation and the results from the
SME-parameter extraction are consistent with those obtained
from the entire data set.  

\section{A Global Solution of Neutrino Oscillations?}
To determine the implications of the allowed SME-parameter values
extracted from the LSND data,
consider the situation where the only nonzero term is $\As\indx$.  This
term was the largest in the maximum-$\ell$ solutions that
allowed for sidereal variation.
In this case, one or more of the SME coefficients,
$(a_L)^X_{\indx}$, $(a_L)^Y_{\indx}$, $(c_L)^{TX}_{\indx}$,
$(c_L)^{TY}_{\indx}$, $(c_L)^{XZ}_{\indx}$ and $(c_L)^{YZ}_{\indx}$
would be nonzero (as can be seen from Eq.~\ref{eq:As}). 

A simple interpretation is that one of the $a_L$-type SME coefficients 
is of order $10^{-19}$~GeV or one of the $c_L$-type is of order $10^{-17}$ 
(or $E\times c_L \sim 10^{-19}$~GeV, where $E$ is the neutrino energy).  
These values would have significant implications in other neutrino 
oscillation experiments and produce effects that have not been 
observed. 
In the simplest class of models, the acceptable maximum scale of Lorentz and
CPT violation for reactor neutrino oscillations is 
$a_L \sim 10^{-21}$~GeV, $c_L \sim 10^{-22}$, and, 
for long-baseline neutrino oscillations, $a_L \sim 10^{-22}$~GeV, 
$c_L \sim 10^{-19}$~\cite{oscillation,Coleman,Pakvasa,Bahcall,Pena-Garay}.
However, there is no theoretical motivation that nature
has chosen a simple solution for neutrino oscillations~\cite{bicycle}.
A global solution of neutrino oscillations with
Lorentz and CPT violation that accommodates
all the data may yet be obtainable within this SME framework.

For this reason, it is important to search for sidereal
variations in other short-baseline neutrino oscillation
experiments. 
The data can be analyzed with the same method as presented here. 

The currently running MiniBooNE experiment~\cite{MiniBooNE}, with a different beam
energy and with a $\nu_\mu$ beam, will be able to test these LSND 
solutions for Lorentz and CPT violating neutrino oscillations 
with a high-statistics appearance measurement.
In particular, a measurement from MiniBooNE would provide an
additional five constraints (Eqs.~\ref{eq:C}-\ref{eq:Bc})
on the SME coefficients.  
The neutrino propagation vectors are different for MiniBooNE as
the neutrino beamline is oriented toward compass north (as opposed 
to east for LSND). Also, the SME coefficients would
be transformed for the neutrino case~\cite{SBL}.
If MiniBooNE collects a significant set of data 
with a $\bar \nu_\mu$ beam, an additional set of constraints 
with antineutrino coefficients would also be obtained.  

Of course, results from other neutrino oscillation experiments
would add further valuable information.  This has been investigated
for Super-Kamiokande~\cite{MMCPT04} and MINOS~\cite{BRCPT04}.

\section{Conclusions}
The neutrino oscillation candidate 
events from the LSND experiment have been examined for 
evidence of sidereal time variation --- a possible signal for 
Lorentz violation in the neutrino sector.
The oscillation excess is consistent with no sidereal time variation. 
An examination of a high-energy subset of the data yields the
same conclusion.

A ``smoking-gun'' for Lorentz violation 
has not been found in the LSND signal.  
However, the data are adequately described
within the SME neutrino 
oscillation formalism that includes both
Lorentz and CPT violation~\cite{oscillation,SBL}.
A maximum-likelihood method was used to determine
allowed parameter regions for SME parameter combinations.
They indicate values on the order of
$10^{-19}$~GeV for $a_L$ and $E \times c_L$. These values are 
in the range expected for Planck-scale effects
in the neutrino sector.  Future results from high-statistics
oscillation experiments will allow more stringent tests of 
the SME framework.

\section{Acknowledgments}
This work was conducted under the auspices of the US Department of Energy,
supported in part by funds provided by the University of California for the 
conduct of discretionary research by Los Alamos National Laboratory.  This
work was also supported by the National Science Foundation.

\end{document}